\documentclass[pre,aps,twocolumn,showpacs,floatfix]{revtex4}

\usepackage{epsfig}
\usepackage{graphicx}
\usepackage{amsmath}

\begin{document}

\title{ Temperature dependent surface relaxation for Al($110$) and 
Mg($10\bar{1}0$) studied by orbital free ab initio molecular dynamics.}

\author{Luis Enrique Gonz\'alez and 
David J. Gonz\'alez}
\affiliation{Departamento de F\'\i sica Te\'orica,
Universidad de Valladolid, 47011 Valladolid, SPAIN.}

\date{\today}

\begin{abstract}
We have performed orbital free ab initio molecular dynamics simulations in order
to study the thermal behaviour of two open surfaces of solid metallic systems, 
namely the ($110$) face of fcc Al and the ($10\bar{1}0$) face of hcp Mg. Our results
reproduce qualitatively both the experimental measurements and previous ab initio
calculations performed with the more costly Kohn-Sham approach of Density
Functional Theory. These calculations can be viewed as a validation test of the 
orbital free method for semiinfinite surfaces, and the results underpin its
reliability.
\end{abstract}

\maketitle

\section{Introduction}

If a bulk metallic crystal at zero temperature is instantaneously separated
into two halves, exposing two pristine surfaces, then the electrons and the 
ions redistribute themselves, responding to the new environment, in order to 
lower the total energy. 
In the simplest cases this results in surface relaxation, 
where the spacing between layers near the surface varies with respect to its 
bulk values \cite{Smolu-Finnis}. 
Screening of the surface and smoothing of the charge usually leads
to an expansion of the first interlayer distance for close-packed faces.
Inner interlayer distances usually change very little for this type of 
surfaces.
For open surfaces, in addition, there is a possibility of diminishing the 
undercoordination felt by the outermost atoms by moving closer to the second
layer, resulting usually in a contraction of the first interlayer distance.
Inner interlayer relaxations also occur, in general leading to a damped 
oscillatory pattern of expansion, contraction, expansion and so on.

When the temperature is raised additional dynamic effects take place and it 
is possible to find experimentally a widely varying thermal behaviour. While 
general theoretical considerations conclude that thermal expansion is 
expected \cite{Jayanthi}, 
some surfaces exhibit an anomalously large effect for its first 
interlayer distance ( Pb(110), Ni(100), Ag(111), Cu(110), Be(0001) )
\cite{Frenken-others-Pohl} 
and some others show a {\em negative} thermal expansion 
coefficient for this first interlayer distance, followed by positive 
coefficients for inner interlayer distances ( Al(110) ) \cite{Goebel} 
or even by an alternating sign behavior ( Mg(10$\bar{1}$0) \cite{Ismail1}, Be(10$\bar{1}$0) \cite{Ismail2}).

In this work we will focus on two of these ``anomalous" systems, namely Al(110)
and Mg(10$\bar{1}$0). Both are simple $sp$ bonded metals, amenable to study 
within a pseudopotential formalism. Moreover, both of them have been studied 
by ab initio methods, namely using the Kohn-Sham (KS) \cite{KS} version of Density 
Functional Theory (DFT) \cite{HK}. For Al(110) ab initio molecular dynamics (AIMD) 
simulations were performed \cite{Marzari}, 
which were able to reproduce both the oscillatory
relaxations and the thermal behaviour observed experimentally. In the case of
Mg(10$\bar{1}$0) theoretical calculations based on the quasiharmonic 
approximation (QHA) and static KS-DFT computations \cite{Ismail1}, 
reproduced also the 
oscillatory pattern both in relaxations and in thermal expansion coefficients.
It should be noted, however, that no AIMD simulations have yet been performed
for this system.

KS-AIMD simulations represent a very powerful tool to study metallic surfaces, 
because, due to the use of DFT, the response of the ions to the rapid 
decrease of the electron density 
near the surface is calculated selfconsistently. However its application has
been very scarce in the literature because of practical difficulties, namely 
they are extremely expensive computationally. Some of this cost can be 
alleviated if one returns to the original formulation of DFT \cite{HK}, 
which uses the
electron density as the only variable in the theory, without any resort to
KS orbitals. In such an orbital free (OF) theory \cite{Madden1}, the 
electronic kinetic
energy must be computed approximately (instead of exactly in the KS version),
and local pseudopotentials are needed (in the KS version also nonlocal 
pseudopotentials can be used).
However, the computing time and memory saved by disposing of the 
orbitals can then 
be invested in studying larger systems for longer times. This, while 
important in solid metallic surfaces, shows its full potential in the study of 
liquid metallic surfaces \cite{GGSs_prl}, 
where the absence of long range order requires the 
use of large samples in order to obtain realistic results.
Note that OF-AIMD still use DFT, and therefore the main power of the AIMD is 
preserved, i.e. the electrons and the ions near the surface respond
selfconsistently to the rearrangement of one another.

While OF-AIMD simulations have been successful in the study of static and
dynamic properties of bulk liquid metals \cite{Madden1,Maddens,GGLSs_method,GGLSs_results}, 
as well as in the understanding of
the thermal properties of some metallic clusters \cite{AguadoLopezs} 
(including an anomalous
variation of the melting temperature in Na clusters with size \cite{Haberland}), 
its more approximate 
character (as compared to KS-AIMD ones) might induce someone to wonder if the 
theory is at all applicable to metallic surfaces, or at least to question 
how accurate it is. In this work we perform OF-AIMD simulations for the two 
solid systems mentioned previously, Al(110) and Mg($10\bar{1}0$), for which the 
KS calculations can be taken as a benchmark. 
The case of Al will permit a direct comparison with both experiment and KS 
simulations, while for Mg the comparison with KS calculations is less direct, 
since no KS simulations were performed; 
nevertheless the comparison with the results obtained within 
the QHA will be valuable anyhow.
We will show that the OF-AIMD method is indeed able to reproduce qualitatively 
the experimental and KS findings, butressing therefore its reliability.

In section \ref{method} we outline the theoretical basis behind the OF-AIMD
simulations. Section \ref{results} shows our results for Al(110) and 
Mg($10\bar{1}0$), and we discuss these results and obtain our conclussions
in section \ref{discuss}.

\section{Method}

\label{method}

A simple $sp$ bonded metal is treated as a set of 
$N$ bare ions with valence $Z$, enclosed in a volume $V$, and 
interacting with $N_{\rm e}=NZ$ 
valence electrons through an electron-ion potential $v(r)$.
The total potential energy of the system can be written, within the 
Born-Oppenheimer approximation, as the sum of the direct ion-ion coulombic 
interaction energy and the ground state energy of the electronic system,   
$E_g$, 
under the external
potential created by the ions, $V_{\rm ext}
(\vec{r},\{\vec{R}_l\}) = \sum_{i=1}^N v(|\vec{r}-\vec{R}_i|)$ ,

\begin{equation}
E(\{\vec{R}_l\}) = \sum_{i<j} \frac{Z^2}{|\vec{R}_i-\vec{R}_j|} +
E_g[\rho_g(\vec{r}),V_{\rm ext}(\vec{r},\{\vec{R}_l\})] \, ,
\end{equation}

\noindent where $\rho_g(\vec{r})$ is the ground state electronic density and 
$\vec{R}_l$ are the ionic positions. 

According to DFT, the ground state electronic 
density, $\rho_g(\vec{r})$,  
can be obtained by minimizing the energy functional
$E[\rho]$, which can be written 

\begin{equation}
E[\rho(\vec{r})] = 
T_s[\rho]+ E_H[\rho]+ E_{\rm xc}[\rho]+ E_{\rm ext}[\rho]
\label{etotal}
\end{equation}

\noindent
where the terms represent, respectively, 
the electronic kinetic energy, $T_s[\rho]$, 
of a non-interacting system of density $\rho(\vec{r})$, 
the classical electrostatic energy (Hartree term), 

\begin{equation}
E_H[\rho] = \frac12 \int \int d\vec{r} \, 
d\vec{s} \, \frac{\rho(\vec{r})\rho(\vec{s})}
{|\vec{r}-\vec{s}|} \, ,
\end{equation}

\noindent
the exchange-correlation
energy, $E_{\rm xc}[\rho]$, for which we adopt the local 
density approximation, and 
finally the electron-ion interaction energy, 
$E_{\rm ext}[\rho]$, where the electron-ion potential has been 
characterized by a local ionic pseudopotential  
which has been constructed within DFT. 

\begin{equation}
E_{\rm ext}[\rho] = \int d\vec{r} \, \rho(\vec{r}) V_{\rm ext}(\vec{r}) \, ,
\end{equation}

In the KS approach to DFT $T_s[\rho]$ is 
calculated exactly by  using single particle orbitals. 
The huge computational effort involved in this approach for large
systems is alleviated in the OF-AIMD approach by use of an explicit but 
approximate functional of the density for $T_s[\rho]$. 
Proposed functionals consist of the von Weizs\"acker term, 

\begin{equation}
T_W[\rho(\vec{r})] = \frac18 \int d\vec{r} \, 
|\nabla \rho(\vec{r})|^2 /\rho(\vec{r}), 
\end{equation}

\noindent 
plus further terms chosen 
in order to reproduce correctly
some exactly known limits. Here, we have used an 
average density model, where 
$T_s=T_W+T_{\beta}$, 

\begin{eqnarray}
T_{\beta} = \frac{3}{10} \int d\vec{r} \, \rho(\vec{r})^{5/3-2\beta}
\tilde{k}(\vec{r})^2 \\
\tilde{k}(\vec{r}) = (2k_F^0)^3 \int d\vec{s} \, k(\vec{s})
w_{\beta}(2k_F^0|\vec{r}-\vec{s}|)   \nonumber
\end{eqnarray}

\noindent
$k(\vec{r})=(3\pi^2)^{1/3} \;  \rho(\vec{r})^{\beta}$, $k_F^0$ is the Fermi 
wavevector for mean electron density $\rho_e = N_e/V$, and $w_{\beta}(x)$ is a 
weight function chosen so that both the linear response theory and 
Thomas-Fermi limits are correctly recovered. Further details  
are given in reference [\onlinecite{GGLSs_method}].

Another key ingredient of the energy functional is the 
the local ion pseudopotential, 
$v_{ps}(r)$,  describing the ion-electron interaction. 
For each system, the $v_{ps}(r)$ has been constructed from first 
principles by fitting, within the same $T_s[\rho]$ functional, the 
displaced valence electronic density induced by an ion 
embedded in a metallic medium as obtained in a KS calculation. 
Further details on the construction of the pseudopotential are given  
in reference  [\onlinecite{GGLSs_method}]. 

Given an ionic configuration, the electronic ground state is obtained, 
the potential energy for the ions is evaluated and the forces acting on them 
found using the Hellmann-Feynman theorem. These are then used to move the ions according to Newton equations of motion into a new configuration, after which
the whole procedure is repeated.

\section{Results}

\label{results}

In order to compare our OF-AIMD data with those obtained by KS calculations we have
used in our simulations exactly the same setup as in the previous KS studies.
As a result, in the case of Al(110) we have  8 layers of 9 atoms each plus a vacuum of $8.5$ \AA \ in our simulation cell, in which the in-plane lattice spacing is 
taken as the experimental one for each temperature considered.
In the case of Mg($10\bar{1}0$) we consider 16 layers of 4 atoms each and an $8.5$ \AA \ vacuum, with again the experimental in-plane lattice spacings.

\subsection{Al($110$)}

The OF-AIMD simulations of Al(110) have been performed for two temperatures close 
to those where experimental data obtained through low energy electron
diffraction (LEED) were reported, namely $T=70$ and $310$ K, and also at a 
higher temperature of $707$ K, near one of the KS-AIMD simulations.

For comparison, the KS-AIMD simulation time was between 5 and 10 ps depending on the 
temperature (for $T=700$ K it was 6 ps) with runs performed on a vector supercomputer (Hitachi S3600), and the OF-AIMD simulation times were between 8 and 16
ps after an initial equilibration time of 2-4 ps 
(8 and 4 ps respectively for $T=707$ K), with runs performed on an Intel Centrino laptop (with a Pentium M processor).

\begin{figure}
\begin{center}
\psfig{file=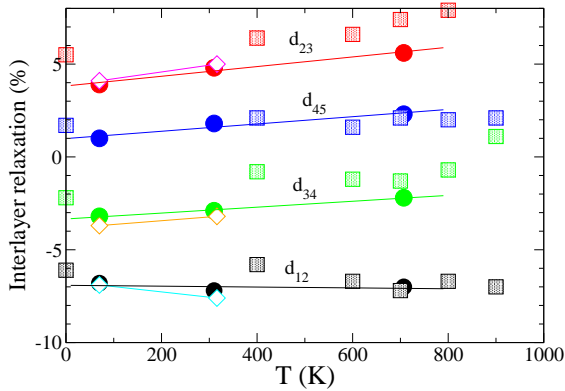,angle=-90,width=90mm}
\end{center}
\caption{Results for the interlayer relaxations in percent 
$100 (d_{ij}(T)-d_{ij}^{\rm bulk}(T))/d_{ij}^{\rm bulk}(T)$, for Al(110).
Full circles: OF-AIMD results. Line: best linear fit to them. 
Hatched squares: uncorrected KS-AIMD results. Lozenges with lines: experimental
data.}
\end{figure}

The interlayer distances obtained from our simulations are plotted in figure 1,
together with the experimental data and the results of the KS-AIMD simulations.
Note that the KS-AIMD results published in reference [\onlinecite{Marzari}] 
were corrected rigidly 
by the difference between a KS calculation at $T=0$ K using the same setup and 
another one using more layers, one atom per layer, and better Brillouin 
zone sampling. 
In figure 1 we have plotted the uncorrected results in order to make a fair 
direct comparison between the OF-AIMD and the KS-AIMD results for the simulation setup used.

\subsection{Mg($10\bar{1}0$)}

The OF-AIMD simulations for this system have been performed near the temperatures
at which the LEED study was performed, which also coincide with the temperatures 
of the KS-QHA calculations, namely $T=106, 308$ and $399$ K. The equilibration and
production runs spanned a time of 4 and 8 ps respectively.

Figure 2 shows the results for this system. For the hcp structure in this orientation
there are two types of interlayer distances, a short interlayer distance, between the
first and second layers, between the third and fourth layers and so on, 
and a long interlayer distance (twice as large in the bulk solid) between the second and third layers, between the fourth and fifth layers and so on.
We have separated figure 2 into two panels in order to appreciate more
clearly the thermal variation of the interlayer distances of both types.

\begin{figure}
\begin{center}
\psfig{file=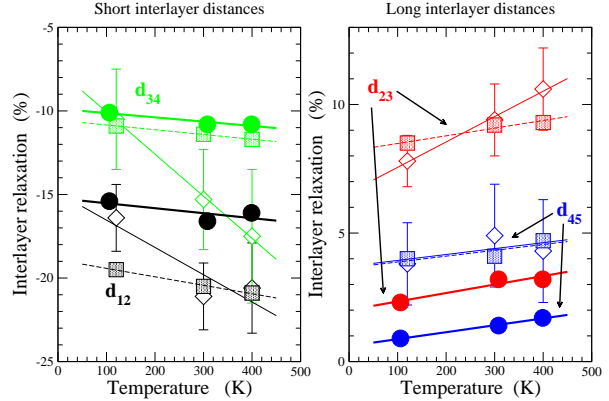,angle=-90,width=90mm}
\end{center}
\caption{Results for the interlayer relaxations in percent 
$100 (d_{ij}(T)-d_{ij}^{\rm bulk}(T))/d_{ij}^{\rm bulk}(T)$, for Mg($10\bar{1}0$).
Full circles: OF-AIMD results. Thick continuous lines: best linear fit to them. 
Hatched squares: KS-QHA results. Dashed lines: best linear fit to them.
Lozenges with error bars: experimental data. Thin continuous lines: best linear fit to them.}
\end{figure}

\section{Discussion and conclusions}

\label{discuss}

Starting with Al(110) we first remark that the OF-AIMD simulations recover
the oscillatory pattern of interlayer relaxation observed experimentally, and also the thermal variation of these data, showing a negative first interlayer thermal 
expansion coefficient and a positive one for all the inner interlayer relaxations.
The agreement with the experimental data at 70 and 310 K is excellent, but we would rather emphasize the reproduction of the trends rather than the numbers.
One of the main reasons for stressing this is the lack of a detailed analysis of size
effects in the simulations. According to recent all-electron first principles 
calculations of the properties of Al surfaces \cite{daSilva}, 
it might be neccessary to include
as many as 23 layers in the simulation for the (110) orientation in order to obtain fully converged results.
This of course needs to be tested for the OF-AIMD simulations before a comparison 
with experimental data can be made at a quantitative level.

Comparing with KS-AIMD results, we find a reasonably good agreement, taking into account the differences in the simulations (kinetic energy functional and pseudopotentials). In any case, again we think that the important point is the reproduction of the trends observed in the KS simulations.

Coming to Mg($10\bar{1}0$), we again remark first that the OF-AIMD results reproduce
the trends of relaxation and thermal variation in this system. The short interlayer
distances both contract and show negative thermal expansion coefficient, while
the long interlayer distances both expand and show positive thermal expansion
coefficient.
The magnitude of the contractions is reproduced with better accuracy than that of 
the expansions, but nevertheless the trend is the correct one.

When comparing with the KS-QHA data we outline three points. First, the magnitude of the relaxations obtained from the KS-QHA is closer to the experimental
one than that of the OF-AIMD simulations. Second, the thermal variation of the 
KS-QHA and OF-AIMD results is very similar, as observed from the slope of the
linear fit to both types of data. And third, both approaches underestimate
largely the thermal variation found in the experimental measurements, which show
a much larger slope for the first three interlayer distances.

Summarizing, the OF-AIMD results for the structure of the open surfaces considered
and their thermal variation reproduce qualitatively the experimental trends. In 
many cases, the results are also very similar to those obtained through more 
expensive methods as KS-DFT, either used within AIMD simulations or within the QHA,
with the only exception of the magnitude of the expansion of the long interlayer
distances in the hcp structure. 
A quantitative comparison with experimental data is at present not sensible, since
a detailed study of size effects in the simulations is necessary, most surely for
Al(110) but probably also in the case of Mg($10\bar{1}0$). This analysis is under 
way and will be presented elsewhere.

In our opinion, the results shown in this work, added to those already
published for bulk metallic liquids and for metallic clusters, further demonstrate 
that the OF-AIMD method is not only practical, but also reliable for the study
of metallic systems, and in particular metallic surfaces.


\begin{references}

\bibitem{Smolu-Finnis} R. Smoluchowski, Phys. Rev. {\bf 60}, 661 (1941);
M. W. Finnis and V. Heine, J. Phys. F: Met. Phys. {\bf 4}, 
L37 (1974)

\bibitem{Jayanthi} C. S. Jayanthi, E. Tossati and A. Fasolino, Phys. Rev. B {\bf 31},
470 (1985); C. S. Jayanthi, E. Tossati and L. Pietronero, Pys. Rev. B {\bf 31}, 3456
(1985)

\bibitem{Frenken-others-Pohl} J. W. M. Frenken, F. Huussen and J. F. van der Veen, Phys. Rev. Lett. {\bf 58}, 401 (1987); Y. Cao and E. Conrad, Phys. Rev. Lett. 
{\bf 65}, 2808 (1990); P. Statiris, H. C. Lu and T. Gustafsson, Phys. Rev. Lett.
{\bf 72}, 3574 (1994); G. Helgesen, D. Gibbs, A. P. Baddorf, D. M. Zehner and 
S. G. J. Mochrie, Phys. Rev. B {\bf 48}, 15320 (1993); K. Pohl, J. H. Cho, K. Terakura, M. Scheffler and E. W. Plummer, Phys. Rev. Lett. {\bf 80}, 2853 (1998).

\bibitem{Goebel} H. G\"obel and P. von Blanckenhagen, Phys. Rev. B {\bf 47}, 2378
(1993)

\bibitem{Ismail1} Ismail, E. W. Plummer, M. Lazzeri and S. de Gironcoli, Phys. Rev.
B {\bf 63}, 233401 (2001)

\bibitem{Ismail2} Ismail, Ph. Hofmann, A. P. Baddorf and E. W. Plummer, Phys. Rev. B
{\bf 66}, 245414 (2002)

\bibitem{KS} W. Kohn and L.J. Sham, Phys. Rev. {\bf 140}, A1133 (1965)

\bibitem{HK} P. Hohenberg and W. Kohn, Phys. Rev. {\bf 136}, 864 (1964)

\bibitem{Marzari} N. Marzari, D. Vanderbilt, A. de Vita and M. C. Payne, Phys. Rev.
Lett. {\bf 82}, 3296 (1999)

\bibitem{Madden1} M. Pearson, E. Smargiassi and P. A. Madden, 
J. Phys.: Condens. Matter \textbf{5}, 3221 (1993)

\bibitem{GGSs_prl} D. J. Gonz\'alez, L. E. Gonz\'alez and M. J. Stott, 
Phys. Rev. Letters {\bf 92}, 085501 (2004);
Phys. Rev. Letters {\bf 94}, 077801 (2005); L. E. Gonz\'alez,
D. J. Gonz\'alez and M. J. Stott, 
J. Chem. Phys. {\bf 123}, 201101 (2005)

\bibitem{Maddens} M. Foley, E. Smargiassi and P. A. Madden, J. Phys: Condens. Matter
{\bf 6}, 5231 (1994); J. A. Anta, B. J. Jesson and P. A. Madden, Phys. Rev. B
{\bf 58}, 6124 (1998); J. A. Anta and P. A. Madden, J. Phys: Condens. Matter 
{\bf 11}, 6099 (1999)

\bibitem{GGLSs_method} L. E. Gonz\'alez, D. J. Gonz\'alez and J. M. L\'opez, 
J. Phys.: Cond. Matter {\bf 13} 7801 (2001);
 D. J. Gonz\'alez, L. E. Gonz\'alez, J. M. L\'opez 
and M. J. Stott, Phys. Rev. B {\bf 65} 184201 (2002) 

\bibitem{GGLSs_results} S. G\'omez, L. E. Gonz\'alez, D. J. Gonz\'alez, M. J. Stott,
S. Dalg\i\c{c} and M. Silbert, J. Non-Cryst. Solids {\bf 250-252}, 163 (1999);
 D. J. Gonz\'alez, L. E. Gonz\'alez, J. M. L\'opez and M. J. Stott, J. Chem. Phys.
{\bf 115}, 2373 (2001); J. Non-Cryst. Solids {\bf 312-314}, 110 (2002);
J. Blanco, D. J. Gonz\'alez, L. E. Gonz\'alez, J. M. L\'opez and M. J. Stott,
J. Non-Cryst. Solids {\bf 312-314}, 148 (2002); Phys. Rev. E {\bf 67} 41204 (2003);
D. J. Gonz\'alez, L. E. Gonz\'alez, J. M. L\'opez and M. J. Stott, EuroPhys. Lett.
{\bf 62}, 42 (2003); Phys. Rev. E {\bf 69}, 31205 (2004); J. Phys: Condens. Matter
{\bf 17}, 1429 (2005)

\bibitem{AguadoLopezs} A. Aguado, J. M. L\'opez, J. A. Alonso and M. J. Stott,
J. Chem. Phys {\bf 111}, 6026 (1999); J. Phys. Chem. B {\bf 105}, 2386 (2001);
A. Aguado, Phys. Rev. B {\bf 63}, 115404 (2001);
A. Aguado, L. M. Molina, J. M. L\'opez and J. A. Alonso, Eur. Phys. J. D {\bf 15},
221 (2001); A. Aguado, L. E. Gonz\'alez and J. M. L\'opez, J. Phys. Chem. B
{\bf 108}, 11722 (2004); A. Aguado and J. M. L\'opez, Phys. Rev. B {\bf 71},
75415 (2005); Phys. Rev. Lett. {\bf 94}, 233401 (2005); 
A. Aguado, J. Phys. Chem. B {\bf 109}, 13043 (2005);
A. Aguado and J. M. L\'opez, Phys. Rev. B {\bf 72}, 205420 (2005)

\bibitem{Haberland} M. Schmidt, R. Kusche, W. KronM\"uller, B. von 
Issendorff and H. Haberland, Phys. Rev. Lett. {\bf 79}, 99 (1997); 
M. Schmidt, R. Kusche, B. von Issendorff and H. Haberland, 
Nature (London) {\bf 393}, 328 (1998)

\bibitem{daSilva} J. L. F. da Silva, Phys. Rev. B {\bf 71}, 195416 (2005)

\end{references}
\end{document}